\documentclass[12pt]{article}
\date{}
\title{Plasma Neutrino Process in Strong Magnetic Field}
\author{{\bf Indranath Bhattacharyya}\\
Department of Mathematics,\\
Government College of Engineering\\
and Ceramic Technology,\\
Government of West Bengal\\
73, Abinash Chandra Banerjee Lane,\\
Kolkata-700010, West Bengal, India.\\
E-mail : $i_{-}bhattacharyya@hotmail.com$\\} \vskip .1in
\begin{document}
\maketitle \baselineskip .3in \noindent
\begin{abstract}
The decay of magneto-plasma into neutrino anti-neutrino pair has
been studied in the framework of the electro-weak interaction
theory. The decay rate is calculated and the expression for the
energy-loss rate is obtained in the extreme relativistic case. The
neutrino luminosity has also been computed for a neutron star. A
comparative study between the decay of magneto-plasma and the
ordinary plasma neutrino process has been outlined in view of the
cooling of highly magnetized star. The effect of this process in
the different regions during the late stages of stellar evolution
is discussed briefly.\vspace{0.5cm}\\\noindent{\bf
Keywords:\hspace{0.2cm}}{Neutrino Emission; Electroweak
interaction; Magnetized Plasma; Decay Rate; Neutron
Stars}\\\noindent{\bf PACS:\hspace{0.2cm}}{13.15.+g; 12.15.-y;
52.20.-j; 75.30.Gw; 97.60.-s; 97.60.Jd }

\end{abstract}
\section{Introduction:} \indent The ordinary plasmon decay into
neutrino anti-neutrino pair is widely discussed topic and it is
believed to be one of the important mechanisms that is responsible
for carrying away the energy from the stellar core. The neutrino
emission from isotropic stellar plasma was considered earlier by
Adams et al. \cite{Adams}. After that it was studied by Braaten
and Segel \cite{Braaten} to do some modifications in the
calculations. It has been shown earlier that the neutrino
bremsstrahlung process is affected in presence of super strong
magnetic field \cite{Bhattacharyya2}. In the same way the decay of
plasma can occur in presence of magnetized environment. In 1969
Canuto et al. \cite{Canuto1, Canuto2} considered the plasma
neutrino process in a strong magnetic field. A consistent study of
the magnetized plasma neutrino process with the inclusion of axial
vector current was carried out by Kennett and Melrose
\cite{Kennett}. Here we have performed a brief study of the
plasmon decay with the influence of magnetic field and its
possible implications in the stars having high core temperature,
density and magnetic field. In this context we have also indicated
the possible region where this process might have some effect.
\section{Calculation of the decay rate and energy loss rate}\indent
The presence of magnetic field produces the anisotropy of plasma
for which it is not possible to divide the electromagnetic waves
into longitudinal and transverse parts unlike the case of
isotropic plasma. In the cold magneto-plasma thermal motion of the
particles are neglected since the phase velocity of such wave is
much larger than the mean thermal velocity. The refractive index
of the cold magneto-plasma is different\cite{Sitenko} from that of
the isotropic plasma. If we ignore the effect of the magnetic
field the refractive index will depend only on the plasma
frequency. \\\indent The matrix element for this process can be
constructed as follows:
$$M_{fi}=-ie\frac{G_{F}}{\sqrt{2}}\varepsilon_{\mu}[C_{V}\Pi^{\mu\rho}(k)+C_{A}\Pi_{A}^{\mu\rho}(k)]
\overline{u}_{\nu}(q_{1})\gamma_{\rho}(1-\gamma_{5})v_{\nu}(q_{2})\eqno{(1)}$$
The term $\Pi^{\mu\rho}(k)$ present in the matrix element
represents the response tensor \cite{Kennett} whereas
$\Pi^{\mu\rho}_{A}(k)$ stands for the same, but associated with
axial vector part. The response tensor depends on the magnetic
field as well as four momentum of the plasmon. Imposing the gauge
invariance restriction one can think that the response tensor
takes the form as
$$\Pi^{\mu\rho}(k)=A(k^{\mu}k^{\rho}-g^{\mu\rho}k^{2})\eqno{(2)}$$
where, $A=A(H)$.\\\indent Similarly we can find the expression for
the axial vector part. Now the decay rate is obtained from the
following expression.
$$\tau=\frac{4\pi\mathcal{S}}{2\omega}\int\Sigma \mid M_{fi}\mid^{2}
\frac{N_{q_{1}}d^{3}q_{1}}{2q_{1}^{0}(2\pi)^{3}}
\frac{N_{q_{2}}d^{3}q_{2}}{2q_{2}^{0}(2\pi)^{3}}
\delta^{4}(k-q_{1}-q_{2})$$
$$=G_{F}^{2}\alpha\frac{k^{2}}{\omega}\mid (g_{V}A+g_{A}A_{5})k^{2} \mid^{2}\eqno{(3)}$$
The term $\mid (g_{V}A+g_{A}A_{5})k^{2} \mid^{2}$ can be obtained
in different approximations. To calculate this term we have used
the result of Kennett and Melrose \cite{Kennett}. If we consider
the relativistic and degenerate plasma the term can be
approximated as
$$\mid(g_{V}A+g_{A}A_{5})k^{2}\mid \approx \frac{\alpha}{3\pi^{4}}(\frac{H}{H_{c}})m_{e}^{4}
[ln(\frac{\mathcal{\mu}}{m_{e}})]\eqno{(4)}$$ Now from the
equations (3) and (4) the decay rate in the extreme relativistic
and highly degenerate case is calculated as follows.
$$\tau\approx 2.198\times10^{-10}(\frac{H}{H_{c}})^{2}\frac{\hbar\omega}{m_{e}c^{2}}(1-\eta_{\ast}^{2})\mid
ln(\frac{\mu}{m_{e}c^{2}})\mid^{2}
\hspace{0.5cm}sec^{-1}\eqno{(5)}$$ Here $\eta_{\ast}$ represents
the refractive index of the plasma that is free from magnetic
influence and so clearly it depends on the plasma frequency but
not on the magnetic field. Finally we obtain an expression for the
energy loss rate as follows:
$$\mathcal{E}\approx 5.74\times10^{3}\times(\frac{H}{H_{c}})^{2}\times
T_{9}^{3}[\frac{1}{3}ln\rho-6.672]\hspace{0.5cm}erg/gm-sec\eqno{(6)}$$
We have computed the neutrino luminosity expressed in the unit of
solar luminosity (Table-1) at the fixed density $\rho=10^{15}$
$gm/cc$ in the temperature range $10^{10} - 10^{11}$ K and
magnetic field $10^{12}- 10^{15}$ G. We have computed the same for
the decay of plasma in absence of magnetic field in the same
table.
\begin{table}
\begin{tabular}{|c|cccc|c|}\hline
$T_{10}$&&&$log\frac{L}{L_{\odot}}$\\\hline &&&With magnetic
field&& Without magnetic field\\\hline & $10^{12}$ & $10^{13}$ &
$10^{14}$ & $10^{15}$ \\\hline
  $1$ & $4.06$ &$6.07$&$8.07$ &$10.07$ & $10.12$\\
  $2$ & $4.97$ & $6.97$ &$8.97$&${\bf 10.97}$&$10.57$\\
  $3$ & $5.50$ & $7.50$ & $9.50$ & ${\bf 11.50}$ & $10.61$\\
  $4$ & $5.87$ & $7.87$ & $9.87$ & ${\bf 11.87}$ & $10.57$\\
  $5$ & $6.16$ & $8.16$ & $10.16$ & ${\bf 12.16}$ & $10.51$\\
  $6$ & $6.40$ & $8.40$ & $10.40$ & ${\bf 12.40}$ & $10.46$\\
  $7$ & $6.60$ & $8.60$ & ${\bf 10.60}$ & ${\bf 12.60}$ & $10.40$\\
  $8$ & $6.78$ & $8.78$ & ${\bf 10.78}$ & ${\bf 12.78}$ & $10.35$\\
  $9$ & $6.93$ & $8.93$ & ${\bf 10.93}$ & ${\bf 12.93}$ & $10.30$\\
  $10$ & $7.07$ & $9.07$ & ${\bf 11.07}$ & ${\bf 13.07}$ & $10.25$\\\hline
\end{tabular}
\caption{Neutrino luminosity  for neutron star($\rho= 10^{15} gm /
cm^{3}$, and magnetic field $H=10^{12}$, $10^{13}$, $10^{14}$,
$10^{15}$ G) due the decay of plasma in presence and absence of
magnetic field respectively in the temperature range $10^{10} -
10^{11}$ K. The bold numbers indicate the former dominates over
the later.}
\end{table}
\section{Discussion:} Adams et al. \cite{Adams} pointed out that the energy
loss rate due to the radiation of neutrino - anti neutrino pairs
by unmagnetized plasmons is very high in the non-relativistic
degenerate region, as much as pair annihilation process.
Eventually our result would not give any significant contribution
here as the magnetic field present here is very low compared to
the critical value. The scenario changes drastically when we
consider this process in the degenerate extreme relativistic
region (e.g. stellar core of a newly born neutron star). From the
Table-1 it is quite clear that in the temperature range $10^{10}-
10^{11}$ K the neutrino luminosity for the decay of magneto-plasma
is very high. When we compare the effect of magneto-plasma decay
with respect to that of unmagnetized plasma, we see in the low
magnetic field of the neutron star the later process is much
dominating the former, but the first one becomes significant with
the increase of magnetic field strength, especially after
exceeding the critical value. From Table-1 it is very much clear
that the presence of magnetic field will have significant effect
in the neutron star when the temperature $\geq 7\times10^{11}$ K
and magnetic field $\geq 10^{14}$ G. It is worth noting that in
case of newly born neutron star having the core temperature more
than $10^{11}$ K the magneto-plasma neutrino process could be more
effective than ordinary plasma neutrino process, even below the
critical magnetic field. Thus the plasma neutrino process in
presence of strong magnetic field plays a crucial role to carry
away the energy from stars in the later stages of the stellar
evolution. It might be one of the very important processes for the
cooling of neutron stars and magnetars.
\section{Acknowledgement:} I am very much thankful to Prof.
{\bf Probhas Raychaudhuri} of The Department of Applied
Mathematics, University of Calcutta, for his continuous help,
suggestions and guidance during preparation of this manuscript.

\end{document}